\documentclass[12pt,11lomcon,cite,epsfig]{article}

\usepackage{epsfig}
\usepackage{amssymb}
\usepackage{amsmath}
\usepackage{cite}
\usepackage{datatool}
\usepackage{forloop}
\usepackage{ifthen}
\usepackage{longtable}
\usepackage{titling}

\makeatletter
\renewcommand{\section}{\@startsection{section}{1}{0pt}{-3.5ex plus -1ex minus -.2ex}{2.3ex plus.2ex}{\centering\Large\bfseries}} \makeatother
\fontsize{14}{14pt}\selectfont
\thanksmarkseries{arabic}
\begin{document}

\title{\textbf{\large Dependence of effective spectrum width of synchrotron radiation on particle
energy}}
\author{V. G. Bagrov  \thanks{Tomsk State University, Russia; Institute of High Current Electronics, SB RAS, Tomsk, Russia; Institute of Physics, University of Sao Paulo, Brazil. E-mail: bagrov@phys.tsu.ru}, D. M. Gitman  \thanks{Tomsk State University, Russia; Institute of Physics, University of Sao Paulo, Brazil;  P.N. Lebedev Physical Institute, Moscow, Russia. E-mail: gitman@if.usp.br }, A. D. Levin \thanks{Institute of Physics, University of Sao Paulo, Brazil.  E-mail:alevin@if.usp.br}, A. S. Loginov \thanks{Tomsk State University, Russia. E-mail: as\_loginov@phys.tsu.ru }, A. D. Saprykin  \thanks{Tomsk State University, Russia. E-mail: alexx.saprykin.phys@gmail.com}}
\maketitle
 
\section*{Abstract}

\quad \quad For an exact quantitative description of spectral properties in
the theory of synchrotron radiation, the concept of effective spectral width
is introduced. In the classical theory, numeric calculations of effective
spectral width (using an effective width not exceeding 100 harmonics) for
polarization components of synchrotron radiation are carried out. The
dependence of the effective spectral width and initial harmonic on the
energy of a radiating particle is established.

\section{Introduction}

\quad \quad As one of the major quantitative characteristics of spectral
distributions for electromagnetic radiation, one commonly uses the concept
of spectral half-width. For spectral distributions having a sharp maximum,
spectral half-width is the most informative physical characteristic.

However, once a spectral distribution has no pronounced maximum, spectral
half-width ceases to be an adequate quantitative characteristic. In
particular, this is exactly the case of spectral distributions for
synchrotron radiation (SR), and therefore SR spectral half-width has neither
been calculated theoretically, nor measured experimentally.

Instead of spectral half-width, the present study proposes to introduce a
new precise quantitative characteristic of SR spectral distributions:
effective spectral width. It is shown how this quantity can be calculated
theoretically, and which physically relevant information can be obtained
using this quantity.

At present, the theory of synchrotron radiation is a fairly well-developed
area of theoretical physics. Its main elements are described in monographs
(e.g. \cite{1,2,3,4,5,6,7}) and numerous articles. Among the most important SR physical
features, one should take into account a high polarization degree of
radiation and a unique structure of spectral distributions. All
theoretically predictable SR properties have been confirmed by experiment.
The development of SR theory allows one not only to predict radiation
characteristics qualitatively, but also to offer exact quantitative
characteristics of physically important properties. For example, the high
degree of SR polarization was qualitatively predicted by theory more than
half a century ago (see, e.g.,\cite{1}); precise quantitative characteristics
for linear polarization were also obtained, whereas such quantitative
characteristics for circular polarization were obtained much later \cite{8,9,10,11}.

In order to set up the problem, we now present some well-known expressions
of the classical SR\ theory for the physical characteristics of synchrotron
radiation, which can be found in \cite{1,2,3,4,5,6,7}.

The spectral-angular distribution for radiation power of SR polarization
components can be written as%
\begin{equation}
W_{s}=W\sum_{\nu =1}^{\infty }\int_{0}^{\pi }f_{s}(\beta ;\,\nu ,\,\theta
)\sin \theta d\theta .  \label{1}
\end{equation}%
Here, the following notation is used: $\theta $ is the angle between the
control magnetic field strength and the radiation field pulse; $\nu $ is the
number of an emitted harmonic; the charge orbital motion rate is $v=c\beta $%
, where $c$ is the speed of light; $W$ is the total radiated power of
unpolarized radiation, which can be written as%
\begin{equation}
W=\frac{2}{3}\frac{ce^{2}}{R^{2}}(\gamma ^{2}-1)^{2}=\frac{%
2e^{4}H^{2}(\gamma ^{2}-1)}{3m_{0}^{2}c^{3}},\ \ \gamma =\frac{1}{\sqrt{%
1-\beta ^{2}}}\,,  \label{2}
\end{equation}%
where $e$ is the particle charge; $R$ is the orbit radius; $H$ is the
control field strength; $m_{0}$ is the charge rest mass; $\gamma $ is the
relativistic factor. The index $s$ numbers the polarization components: $s=2$
corresponds to the $\sigma $-component of linear polarization; $s=3$
corresponds to the $\pi $-component of linear polarization; $s=1$
corresponds to right-hand circular polarization; $s=-1$ corresponds to
left-hand circular polarization; $s=0$ corresponds to the power of
unpolarized radiation. The functions $f_{s}(\beta ;\,\nu ,\,\theta )$ have
the form  \cite{1,2,3,4,5,6,7}%
\begin{equation}
f_{2}(\beta ;\,\nu ,\,\theta )=\frac{3\nu ^{2}}{2\gamma ^{4}}J_{\nu
}^{\prime 2}(x);\ \ f_{3}(\beta ;\,\nu ,\,\theta )=\frac{3\nu ^{2}}{2\gamma
^{4}}\frac{\cos ^{2}\theta }{\beta ^{2}\sin ^{2}\theta }J_{\nu }^{2}(x); 
\notag
\end{equation}%
\begin{equation}
f_{\pm 1}(\beta ;\,\nu ,\,\theta )=\frac{3\nu ^{2}}{4\gamma ^{4}}\left[
J_{\nu }^{\prime }(x)\pm \varepsilon \frac{\cos \theta }{\beta \sin \theta }%
J_{\nu }(x)\right] ^{2};\ \ x=\nu \beta \sin \theta ;\ \ \varepsilon =-\frac{%
e}{|e|};  \notag
\end{equation}%
\begin{equation}
f_{0}(\beta ;\,\nu ,\,\theta )=f_{2}(\beta ;\,\nu ,\,\theta )+f_{3}(\beta
;\,\nu ,\,\theta )=f_{1}(\beta ;\,\nu ,\,\theta )+f_{-1}(\beta ;\,\nu
,\,\theta )\,.  \label{3}
\end{equation}%
Here, $J_{\nu }(x)$ and $J_{\nu }^{\prime }(x)$ are the Bessel functions and
their derivatives. In what follows, the case of an electron is considered,
which corresponds to $\varepsilon =1$.

\section{Spectral distribution for polarization components of synchrotron
radiation in the upper half-space}

\quad \quad It is well known \cite{1,2,3,4,5,6,7} that the angle range $0\leqslant \theta
<\pi /2$ (this range will be called the upper half-space) is dominated by
right-hand circular polarization, and the angle range $\pi /2<\theta
\leqslant \pi $ (this range will be called the lower half-space) is
dominated by left-hand circular polarization (exact quantitative
characteristics of SR properties were first obtained in \cite{8,9,10,11}). However,
if we integrate in (\ref{1}) over $\theta \ (0\leqslant \theta \leqslant \pi
)$, then the differences in the spectral distribution of right- and
left-hand circular polarizations disappear. To reveal these differences, the
expressions (\ref{1}) can be represented as%
\begin{equation}
W_{s}=W\left[ \Phi _{s}^{(+)}(\beta )+\Phi _{s}^{(-)}(\beta )\right] \,;\ \
\Phi _{s}^{(+)}(\beta )=\sum_{\nu =1}^{\infty }F_{s}^{(+)}(\beta ;\,\nu
)\,,\ \ \Phi _{s}^{(-)}(\beta )=\sum_{\nu =1}^{\infty }F_{s}^{(-)}(\beta
;\,\nu )\,;  \notag
\end{equation}%
\begin{equation}
F_{s}^{(+)}(\beta ;\,\nu )=\int_{0}^{\pi /2}f_{s}(\beta ;\,\nu ,\,\theta
)\sin \theta d\theta \,,\ \ F_{s}^{(-)}(\beta ;\,\nu )=\int_{\pi /2}^{\pi
}f_{s}(\beta ;\,\nu ,\,\theta )\sin \theta d\theta \,,  \label{4}
\end{equation}%
and it suffices to study the properties of functions $F_{s}^{(+)}(\beta
;\,\nu )$ (respectively, the properties of functions $\Phi _{s}^{(+)}(\beta
) $), due to the evident relations%
\begin{equation}
F_{s}^{(-)}(\beta ;\,\nu )=F_{s}^{(+)}(\beta ;\,\nu )\,,\ \ \Phi
_{s}^{(-)}(\beta )=\Phi _{s}^{(+)}(\beta )\,,\ \ s=0,\,2,\,3;  \notag
\end{equation}%
\begin{equation}
F_{1}^{(-)}(\beta ;\,\nu )=F_{-1}^{(+)}(\beta ;\,\nu )\,,\ \ \Phi
_{1}^{(-)}(\beta )=\Phi _{-1}^{(+)}(\beta )\,,  \notag
\end{equation}%
\begin{equation}
F_{-1}^{(-)}(\beta ;\,\nu )=F_{1}^{(+)}(\beta ;\,\nu )\,,\ \ \Phi
_{-1}^{(-)}(\beta )=\Phi _{1}^{(+)}(\beta )\,.  \label{5}
\end{equation}

Integration over $\theta $ in the upper half-space $0\leqslant \theta
\leqslant \pi /2$ in (\ref{4}) can be carried out exactly, yielding the
expressions%
\begin{equation}
F_{2}^{(+)}(\beta ;\,\nu )=\frac{3\nu }{4\,\gamma \,^{4}\beta ^{3}}\left[
2\beta ^{2}J_{2\nu }^{\prime 2}\int_{0}^{2\nu \beta }J_{2\nu }(x)dx-2\nu
\beta \int_{0}^{2\nu \beta }\frac{J_{2\nu }(x)}{x}dx\right] \,,  \notag
\end{equation}%
\begin{equation}
F_{3}^{(+)}(\beta ;\,\nu )=\frac{3\nu }{4\,\gamma \,^{4}\beta ^{3}}\left[
2\nu \beta \int_{0}^{2\nu \beta }\frac{J_{2\nu }(x)}{x}dx-\int_{0}^{2\nu
\beta }J_{2\nu }(x)dx\right] \,,  \notag
\end{equation}%
\begin{equation}
F_{0}^{(+)}(\beta ;\,\nu )=F_{2}^{(+)}(\beta ;\,\nu )+F_{3}^{(+)}(\beta
;\,\nu )=\frac{3\nu }{4\,\gamma \,^{4}\beta ^{3}}\left[ 2\beta ^{2}J_{2\nu
}^{\prime 2})\int_{0}^{2\nu \beta }J_{2\nu }(x)dx\right] \,,  \notag
\end{equation}%
\begin{equation}
F_{\pm 1}^{(+)}(\beta ;\,\nu )=\frac{1}{2}F_{0}^{(+)}(\beta ;\,\nu )\pm 
\frac{3\nu J\,_{\nu }^{2}(\nu \beta )}{4\,\gamma \,^{4}\beta ^{2}}\,.
\label{6}
\end{equation}

The sums over the harmonics $\nu $ in (\ref{4}) can also be calculated
exactly, yielding the expressions 
\begin{equation}
\Phi _{2}^{(+)}(\beta )=\frac{6+\beta ^{2}}{16}\,,\ \ \Phi _{3}^{(+)}(\beta
)=\frac{2-\beta ^{2}}{16}\,,\ \ \Phi _{0}^{(+)}(\beta )=\frac{1}{2}\,,\ \
\Phi _{\pm 1}^{(+)}(\beta )=\frac{1}{4}\left[ 1\pm \frac{3}{4}\chi
_{1}(\beta )\right] \,.  \label{7}
\end{equation}%
The function $\chi _{1}(\beta )$ introduced above was defined and studied in
\cite{8}. In particular, it was shown \cite{8} that in the segment $0\leqslant \beta
\leqslant 1$ the function $\chi _{1}(\beta )$ is finite and decreases
monotonously; at the end of this segment, it takes the following values:%
\begin{equation}
\chi _{1}(0)=1\,,\ \ \chi _{1}(1)=\frac{4}{\pi \sqrt{3}}\,.  \label{8}
\end{equation}

\section{Effective spectral width for polarization components of
synchrotron radiation}

\quad \quad As one of the quantitative characteristics of physical
properties for spectral distributions of SR polarization components, it is
proposed to introduce the concept of effective spectral width $\Lambda
_{s}(\beta )$. Let us define $\Lambda _{s}(\beta )$ as follows.

For each fixed value of $\beta $, we examine the quantities%
\begin{equation}
\widetilde{\Phi }_{s}(\beta ;\,\nu ^{(1)},\,\nu ^{(2)})=\sum_{\nu =\nu
^{(1)}}^{\nu ^{(2)}}F_{s}^{(+)}(\beta ;\,\nu )\,,\ \ 1\leqslant \nu
^{(1)}\leqslant \nu ^{(2)}\leqslant \infty \,.  \label{9}
\end{equation}%
The following relations are obvious:%
\begin{equation}
\Phi _{s}^{(+)}(\beta )=\widetilde{\Phi }_{s}(\beta ;\,\nu ^{(1)}=1,\,\nu
^{(2)}=\infty )\,;\ \ \Phi _{s}^{(+)}(\beta )>\widetilde{\Phi }_{s}(\beta
;\,\nu ^{(1)}\geqslant 1,\,\nu ^{(2)}<\infty )\,.  \label{10}
\end{equation}%
Consider the set of values $\nu ^{(1)},\,\,\nu ^{(2)}\,(1\leqslant \nu
^{(1)}\leqslant \nu ^{(2)}<\infty )$ such that satisfy the inequality 
\begin{equation}
\widetilde{\Phi }_{s}(\beta ;\,\nu ^{(1)},\,\nu ^{(2)})\geqslant \frac{1}{2}%
\Phi _{s}^{(+)}(\beta )\,.  \label{11}
\end{equation}%
Obviously, such values $\nu ^{(1)},\,\nu ^{(2)}$ do exist for any $\beta $
(for example, $\nu ^{(1)}=1$ necessarily yields such a finite value $\nu
^{(2)}$). It is equally obvious that condition (\ref{11}) alone is generally
insufficient to determine the pair of values $\nu ^{(1)},\,\nu ^{(2)}$. Let
us now choose such $\nu _{s}^{(1)}(\beta )$, $\,\nu _{s}^{(2)}(\beta )$ that
the condition (\ref{11}) should provide the minimum of the difference $\nu
_{s}^{(2)}(\beta )-\nu _{s}^{(1)}(\beta )$, as well as the minimum of the
non-negative value%
\begin{equation}
\widetilde{\Phi }_{s}(\beta ;\,\nu _{s}^{(1)}(\beta ),\,\nu _{s}^{(2)}(\beta
))-\frac{1}{2}\Phi _{s}^{(+)}(\beta )\geqslant 0\,.  \label{12}
\end{equation}%
The effective spectral width $\Lambda _{s}(\beta )$ is defined by the
expression%
\begin{equation}
\Lambda _{s}(\beta )=\nu _{s}^{(2)}(\beta )-\nu _{s}^{(1)}(\beta )+1,\ \
\Lambda _{s}(\beta )\geqslant 1\,.  \label{13}
\end{equation}

Consequently, effective spectral width is the minimum spectral range that
accounts for at least half of all the radiated power of a given polarization
component. The harmonics $\nu _{s}^{(1)}(\beta )$ and $\nu _{s}^{(2)}(\beta
) $ determine the beginning and the end of this minimum spectral range.

A definition equivalent to the one presented above for effective spectral
width can be given using the concept of partial contributions $P_{s}(\beta
;\nu )$ for individual spectral harmonics, introduced in \cite{12}. Namely, we
suppose, according to \cite{12},\textbf{\ }%
\begin{equation}
P_{s}(\beta ;\nu )=\frac{F_{s}^{(+)}(\beta ;\,\nu )}{\Phi _{s}^{(+)}(\beta )}%
\,.  \label{14}
\end{equation}%
Then (\ref{4}) implies the property%
\begin{equation}
\sum_{\nu =1}^{\infty }P_{s}(\beta ;\nu )=1\,.  \label{15}
\end{equation}%
We choose some values $\nu _{s}^{(1)}(\beta )$ and $\nu _{s}^{(2)}(\beta )$
such that the minimum difference $\nu _{s}^{(2)}(\beta )-\nu
_{s}^{(1)}(\beta )$ should provide the minimum of the non-negative value%
\begin{equation}
\sum_{\nu =\nu _{s}^{(1)}(\beta )}^{\nu _{s}^{(2)}(\beta )}P_{s}(\beta ;\nu
)-\frac{1}{2}\geqslant 0\,.  \label{16}
\end{equation}%
Introducing $\Lambda _{s}(\beta )$ in accordance with (\ref{13}), we arrive
at the following equivalent definition: effective spectral width is the
minimum spectral range at which the sum of partial contributions for
individual harmonics is not less than $1/2$.

In practice, the most interesting case is the ultra-relativistic limit ($%
\beta \approx 1$, equivalent to $\gamma \gg 1$). In this case, the
analytical study of effective spectral width and other physically
interesting quantitative characteristics for spectral distributions of SR
polarization components can be significantly extended. This study was
carried out in \cite{13}.

Given a particular value of $\beta $ (or $\gamma $), it is a purely
computational task to obtain the exact values of $\Lambda _{s}(\beta )$ and $%
\nu _{s}^{(1)}(\beta )$. In this article, we present a numerical study of
the region $1\leqslant \Lambda _{s}(\beta )\leqslant 100$. It is essential
to observe the following. The effective width $\Lambda _{s}(\beta )$ is a
positive integer, so there exists a range of $\beta $ (corresponding to a
range of $\gamma $; hereinafter, we only indicate $\gamma $) in which $%
\Lambda _{s}(\beta )$ is constant.

\section{Analysis of numerical results for effective spectral width of
synchrotron radiation}

\quad \quad The main results of a numerical study for effective spectral
width of SR polarization components are given by our Table.

The numerical study is carried out as follows. For each type of polarization 
$s$, we examine the sequences of integers $\Lambda _{s}=1\,,2\,,3\,...$ and $%
\nu _{s}^{(1)}=1\,,2\,,3\,...$ (it is evident that $\nu _{s}^{(2)}=\nu
_{s}^{(1)}+\Lambda _{s}-1$) and determine the regions of values $\gamma _{s}$
for which the condition (\ref{16}) (equivalent to (\ref{12})) is satisfied.
It is clear that the boundary points of possible regions for $\gamma _{s}$
can be found, according to (\ref{16}), as solutions of the equations%
\begin{equation}
\sum_{l=0}^{\Lambda _{s}-1}P_{s}(\beta ;\nu _{s}^{(1)}+l)-\frac{1}{2}=0\,.
\label{17}
\end{equation}%
The roots $\beta _{s}=\beta _{s}(\Lambda _{s},\nu _{s}^{(1)})$ of these
equations determine the boundary points $\gamma _{s}=\gamma _{s}(\Lambda
_{s},\nu _{s}^{(1)})$, in accordance with (\ref{2}).

Let us consider in more detail the calculation method and the results
pertaining to the $\sigma $- component of linear polarization. These results
are given in the column $s=2$ of our Table. We denote $\nu
_{2}^{(1)}=k=1\,,2\,,3\,...$.

At the first step, we examine the smallest possible value $\Lambda _{2}=1$.
In this case, equations (\ref{17}) have the form%
\begin{equation}
P_{2}(\beta ;k)-\frac{1}{2}=0\,.  \label{18}
\end{equation}%
According to the results of [12], all partial contributions at $k\geqslant 2$
are such that $P_{2}(\beta ;k)<1/2$ for all the values of $\gamma $, whereas
in the first harmonic ($k=1$) such a region of values $\gamma $ does exist,
and equation (\ref{18}) at $k=1$ has a single root, $\gamma _{2}=\gamma
_{2}(1\,,1)$, shown in the Table.

Consequently, in the region of values $\gamma $%
\begin{equation}
1\leqslant \gamma \leqslant \gamma _{2}(1\,,1),  \label{19}
\end{equation}%
for the $\sigma $-component of linear polarization, the effective spectral
width is $\Lambda =1$, where $\nu _{2}^{(1)}=1$. There are no other values
of $\tilde{\nu}_{2}^{(1)}$ and $\tilde{\gamma}_{2}$ for $\Lambda =1$, which
is also indicated in the Table.

Next, we examine the value $\Lambda _{2}=2$. In this case, equations (\ref%
{17}) have the form%
\begin{equation}
P_{2}(\beta ;k)+P_{2}(\beta ;k+1)-\frac{1}{2}=0\,.  \label{20}
\end{equation}%
Since at the previous step it has been established that in the region $%
1\leqslant \gamma \leqslant \gamma _{2}(1\,,1)$ the effective width is $%
\Lambda _{2}=1$, it is required to examine the solutions of equations (\ref%
{20}) only in the region%
\begin{equation}
\gamma >\gamma _{2}(1\,,1)\,,  \label{21}
\end{equation}%
Analysis of equations (\ref{20}) shows that these equations have a unique
solution, $\gamma _{2}(2\,,1)$, only for $k=1$, where $\gamma
_{2}(1\,,1)<\gamma _{2}(2\,,1)$.

Consequently, in the region of values $\gamma $%
\begin{equation}
\gamma _{2}(1\,,1)<\gamma \leqslant \gamma _{2}(2\,,1)\,,  \label{22}
\end{equation}%
the effective width is $\Lambda _{2}=2$, where $\nu _{2}^{(1)}=1$. There are
no other values of $\tilde{\nu}_{2}^{(1)}$ and $\tilde{\gamma}_{2}$ for $%
\Lambda =2$.

At the next step, we examine the value $\Lambda _{2}=3$. Equations (\ref{17}%
) at $\Lambda _{2}=3$ have the form 
\begin{equation}
P_{2}(\beta ;k)+P_{2}(\beta ;k+1)+P_{2}(\beta ;k+2)-\frac{1}{2}=0\,,
\label{23}
\end{equation}%
and we should only be concerned with solutions of these equations that
belong to the region%
\begin{equation}
\gamma >\gamma _{2}(2\,,1)\,,  \label{24}
\end{equation}%
since at the previous steps it has been found that the effective spectral
width at $1\leqslant \gamma \leqslant \gamma _{2}(2\,,1)$ is such that $%
\Lambda _{2}\leqslant 2$. It turns out that under restriction (\ref{24})
equations (\ref{23}) possess the solutions $\gamma _{2}(3\,,1)$ at $k=1$,
and $\gamma _{2}(3\,,2)$ at $k=2$. These solutions obey the inequalities $%
\gamma _{2}(2\,,1)<\gamma _{2}(3\,,2)<\gamma _{2}(3\,,1)$.

Consequently, in the region of values $\gamma $%
\begin{equation}
\gamma _{2}(2\,,1)<\gamma \leqslant \gamma _{2}(3\,,1)\,,  \label{25}
\end{equation}%
the effective width is $\Lambda _{2}=3$, where $\nu _{2}^{(1)}=1$. However,
inside the region (\ref{25}) there is a subregion:%
\begin{equation}
\gamma _{2}(2\,,1)<\gamma \leqslant \gamma _{2}(3\,,2)<\gamma _{2}(3\,,1)\,,
\label{26}
\end{equation}%
which admits a larger value, $\nu _{2}^{(1)}=\tilde{\nu}_{2}=2$ , as shown
in the Table.

At the next step, we examine $\Lambda _{2}=4\,,5\,,6$. For these values of $%
\Lambda _{2}$ the results of calculation are similar to those obtained for $%
\Lambda _{2}=3$, namely, these are the values of $\gamma $ that belong to
the region 
\begin{equation}
\gamma _{2}(\Lambda _{2}-1\,,1)<\gamma \leqslant \gamma _{2}(\Lambda
_{2}\,,1)\,,  \label{27}
\end{equation}%
with the effective width being equal to the corresponding $\Lambda _{2}$,
where $\nu _{2}^{(1)}=1$. However, inside the region (\ref{27}) there is a
subregion: 
\begin{equation}
\gamma _{2}(\Lambda _{2}-1\,,1)<\gamma \leqslant \gamma _{2}(\Lambda
_{2}\,,2)<\gamma _{2}(\Lambda _{2}\,,1)\,,  \label{28}
\end{equation}%
which admits a larger value, $\nu _{2}^{(1)}=\tilde{\nu}_{2}=2$, as shown in
the Table.

At the next steps, we examine $\Lambda _{2}=7\div 11$, and each step to
follow involves solutions of equations (\ref{17}) only in the region of
values $\gamma $ being larger than those at the previous step. For these
values $\Lambda _{2}$, equations (\ref{17}) possess the solutions $\gamma
_{2}(\Lambda _{2},\nu _{2}^{(1)})$ corresponding to the three values $\nu
_{2}^{(1)}=1$, $\nu _{2}^{(1)}=2$, $\nu _{2}^{(1)}=\tilde{\nu}_{2}=3$, and
satisfying the inequalities 
\begin{equation}
\gamma _{2}(\Lambda _{2}-1,\nu _{2}^{(1)})<\gamma _{2}(\Lambda
_{2}\,,k)<\gamma _{2}(\Lambda _{2}\,,n)\,,\ \ k>n.  \label{29}
\end{equation}%
In the Table, we indicate the smallest value $\nu _{2}^{(1)}=1$ and the
respective largest value $\gamma _{2}(\Lambda _{2},1)$, as well as the
largest value $\nu _{2}^{(1)}=\tilde{\nu}_{2}=3$ and the respective smallest
value $\gamma _{2}(\Lambda _{2},3)$. For $\Lambda _{2}=7\div 11$, the
corresponding regions of $\gamma $ are determined by relation (\ref{27}),
but now there are two subregions: one is determined by relation (\ref{28}),
so that the values of $\gamma $ admit $\nu _{2}^{(1)}=2$; the other one is
determined by the condition 
\begin{equation}
\gamma _{2}(\Lambda _{2}-1\,,1)<\gamma \leqslant \gamma _{2}(\Lambda
_{2}\,,3)<\gamma _{2}(\Lambda _{2}\,,2)<\gamma _{2}(\Lambda _{2}\,,1)\,,
\label{30}
\end{equation}%
and this region of $\gamma $ also admits the value $\nu _{2}^{(1)}=3$.

The next step concerns the region $\Lambda _{2}=12\div 27$. Here, it is
essential that the smallest possible value is $\nu _{2}^{(1)}=2$; otherwise,
the results are identical to the case $\Lambda _{2}=7\div 11$.

In general, the column $s=2$ of the Table indicates for each $\Lambda _{2}$
the smallest possible value $\nu _{2}^{(1)}=\nu _{2}^{(1)}(\Lambda _{2})$
and the corresponding largest value $\gamma _{2}=\gamma _{2}(\Lambda
_{2},\nu _{2}^{(1)}(\Lambda _{2}))$, as well as the largest possible value $%
\tilde{\nu}_{2}$ for this $\Lambda _{2}$ and the corresponding smallest
value $\tilde{\gamma}_{2}$. Possible intermediate values $\nu _{2}^{(1)}$
between the smallest $\nu _{2}^{(1)}(\Lambda _{2})$ and the largest\ $\tilde{%
\nu}_{2}$, as well as the respective intermediate values $\gamma _{2}$, are
not specified. The intermediate values of $\gamma _{2}$ always satisfy
relations (\ref{29}).

Consequently, in the column $s=2$ we indicate the regions of values $\gamma $
\begin{equation}
\gamma _{2}(\Lambda _{2}-1,\nu _{2}^{(1)}(\Lambda _{2}-1))<\gamma \leqslant
\gamma _{2}(\Lambda _{2},\nu _{2}^{(1)}(\Lambda _{2})),  \label{31}
\end{equation}%
for which the effective spectral width for the $\sigma $-component of SR
linear polarization equals to $\Lambda _{2}$.\ We also indicate the initial
points of the effective spectral width. These points are not determined
uniquely. For the smallest initial value $\nu _{2}^{(1)}$, the range of
values (\ref{31}) taken by $\gamma $ is the largest one, while this region
is the smallest one for the largest possible value $\tilde{\nu}_{2}.$

For the other polarization components, the results of calculation are given
in the respective columns of the Table. In particular, the Table shows that
for equal values $\Lambda _{s}$ the corresponding values $\gamma _{s}$ obey
the inequalities 
\begin{equation}
\gamma _{3}>\gamma _{1}>\gamma _{0}>\gamma _{2}>\gamma _{-1}\,.  \label{32}
\end{equation}%
At a fixed energy $\gamma $, the corresponding values of $\Lambda _{s}$ are
restricted by 
\begin{equation}
\Lambda _{3}<\Lambda _{1}<\Lambda _{0}<\Lambda _{2}<\Lambda _{-1}\,.
\label{33}
\end{equation}

In this way, for each polarization component of synchrotron radiation we
have found energy regions at which the effective spectral width equals to $%
\Lambda _{s}$, and the initial harmonic of this effective width is
determined. Numeric calculations have been carried out in the case $\Lambda
_{s}\leqslant 100$.

In the ultrarelativistic case, the corresponding results have been obtained
in \cite{13}.

\DTLsettabseparator
\begin{filecontents*}{table_out.dat}
1	1	1.1434	1	1.1434	1	1.2363	1	1.2363	1	1.1592	1	1.1592	1	1.1062	1	1.1062	1	1.1712	1	1.1712
2	1	1.2955	1	1.2955	1	1.4519	1	1.4519	1	1.3204	1	1.3204	1	1.2348	1	1.2348	1	1.3411	1	1.3411
3	1	1.4179	2	1.3237	1	1.6126	1	1.6126	1	1.4476	1	1.4476	1	1.3440	2	1.3233	1	1.4737	1	1.4737
4	1	1.5215	2	1.4740	1	1.7435	1	1.7435	1	1.5543	2	1.4831	2	1.4394	2	1.4394	1	1.5843	2	1.4987
5	1	1.6121	2	1.5827	1	1.8554	1	1.8554	1	1.6472	2	1.5990	2	1.5323	3	1.4808	1	1.6802	2	1.6210
6	1	1.6932	2	1.6741	1	1.9540	2	1.8592	1	1.7301	2	1.6949	2	1.6125	3	1.5854	1	1.7657	2	1.7211
7	1	1.7669	3	1.6998	1	2.0428	2	1.9638	1	1.8053	2	1.7786	2	1.6841	3	1.6686	1	1.8431	2	1.8082
8	1	1.8347	3	1.7860	1	2.1238	2	2.0562	1	1.8744	2	1.8537	2	1.7493	4	1.7015	1	1.9142	2	1.8860
9	1	1.8977	3	1.8610	1	2.1986	2	2.1394	1	1.9386	3	1.8770	2	1.8094	4	1.7786	1	1.9800	2	1.9570
10	1	1.9566	3	1.9286	1	2.2680	2	2.2161	1	1.9986	3	1.9482	2	1.8655	4	1.8455	1	2.0416	2	2.0225
11	1	2.0120	3	1.9906	1	2.3329	2	2.2868	1	2.0550	3	2.0131	3	1.9197	5	1.8721	1	2.0995	3	2.0448
12	2	2.0657	4	2.0129	1	2.3949	2	2.3537	1	2.1084	3	2.0731	3	1.9711	5	1.9353	1	2.1542	3	2.1072
13	2	2.1167	4	2.0723	1	2.4532	2	2.4165	1	2.1591	3	2.1292	3	2.0196	5	1.9922	1	2.2061	3	2.1654
14	2	2.1652	4	2.1273	1	2.5097	2	2.4751	1	2.2074	3	2.1819	3	2.0656	5	2.0445	1	2.2557	3	2.2200
15	2	2.2114	4	2.1788	1	2.5628	2	2.5314	1	2.2537	3	2.2318	3	2.1094	6	2.0690	1	2.3030	3	2.2716
16	2	2.2559	4	2.2275	1	2.6140	2	2.5851	1	2.2980	3	2.2793	3	2.1512	6	2.1190	1	2.3485	3	2.3207
17	2	2.2984	4	2.2737	1	2.6623	2	2.6363	1	2.3407	4	2.2999	3	2.1913	6	2.1656	1	2.3921	3	2.3675
18	2	2.3392	4	2.3177	1	2.7107	2	2.6846	1	2.3818	4	2.3457	4	2.2302	6	2.2094	1	2.4342	3	2.4123
19	2	2.3787	4	2.3598	1	2.7557	2	2.7324	1	2.4215	4	2.3894	4	2.2682	7	2.2315	1	2.4748	3	2.4553
20	2	2.4168	5	2.3807	1	2.7998	2	2.7779	2	2.4603	4	2.4313	4	2.3049	7	2.2738	1	2.5141	3	2.4967
21	2	2.4538	5	2.4215	1	2.8422	2	2.8217	2	2.4980	4	2.4716	4	2.3402	7	2.3139	1	2.5522	4	2.5154
22	2	2.4895	5	2.4608	1	2.8831	2	2.8636	2	2.5345	4	2.5104	4	2.3744	7	2.3521	1	2.5891	4	2.5556
23	2	2.5243	5	2.4985	1	2.9225	2	2.9035	2	2.5700	4	2.5479	4	2.4076	7	2.3886	1	2.6250	4	2.5943
24	2	2.5581	5	2.5350	1	2.9619	2	2.9440	2	2.6045	4	2.5842	4	2.4397	8	2.4091	1	2.6599	4	2.6318
25	2	2.5910	5	2.5702	1	2.9991	2	2.9827	2	2.6380	4	2.6194	4	2.4710	8	2.4446	1	2.6938	4	2.6681
26	2	2.6231	5	2.6043	1	3.0356	2	3.0201	2	2.6707	4	2.6536	5	2.5017	8	2.4787	1	2.7269	4	2.7033
27	2	2.6545	5	2.6374	1	3.0715	2	3.0566	2	2.7025	5	2.6712	5	2.5318	8	2.5116	2	2.7596	4	2.7376
28	3	2.6849	6	2.6558	1	3.1063	2	3.0922	2	2.7336	5	2.7045	5	2.5611	8	2.5434	2	2.7914	4	2.7709
29	3	2.7148	6	2.6880	1	3.1401	2	3.1266	2	2.7640	5	2.7369	5	2.5897	9	2.5623	2	2.8226	4	2.8033
30	3	2.7443	6	2.7193	1	3.1728	3	3.1407	2	2.7937	5	2.7685	5	2.6176	9	2.5933	2	2.8530	4	2.8350
31	3	2.7731	6	2.7499	1	3.2053	3	3.1740	2	2.8227	5	2.7992	5	2.6449	9	2.6233	2	2.8828	4	2.8659
32	3	2.8012	6	2.7797	1	3.2374	3	3.2078	2	2.8512	5	2.8292	5	2.6716	9	2.6525	2	2.9119	4	2.8960
33	3	2.8288	6	2.8088	1	3.2690	3	3.2415	2	2.8790	5	2.8585	5	2.6977	9	2.6808	2	2.9404	4	2.9256
34	3	2.8559	6	2.8373	1	3.2995	3	3.2728	2	2.9064	5	2.8872	6	2.7234	10	2.6983	2	2.9684	5	2.9414
35	3	2.8824	6	2.8652	1	3.3297	3	3.3039	2	2.9331	5	2.9152	6	2.7487	10	2.7260	2	2.9959	5	2.9703
36	3	2.9085	6	2.8926	1	3.3592	3	3.3346	2	2.9594	5	2.9427	6	2.7736	10	2.7530	2	3.0228	5	2.9986
37	3	2.9339	7	2.9086	1	3.3885	3	3.3646	2	2.9853	5	2.9696	6	2.7979	10	2.7793	2	3.0492	5	3.0263
38	3	2.9589	7	2.9352	1	3.4186	3	3.3946	2	3.0106	5	2.9960	6	2.8219	10	2.8050	2	3.0752	5	3.0535
39	3	2.9835	7	2.9612	1	3.4461	3	3.4241	2	3.0356	6	3.0112	6	2.8453	10	2.8302	2	3.1007	5	3.0801
40	3	3.0078	7	2.9869	1	3.4723	3	3.4511	2	3.0601	6	3.0371	6	2.8684	11	2.8465	2	3.1258	5	3.1063
41	3	3.0316	7	3.0121	1	3.4998	3	3.4793	2	3.0842	6	3.0625	6	2.8911	11	2.8712	2	3.1505	5	3.1319
42	3	3.0550	7	3.0367	1	3.5270	3	3.5074	2	3.1079	6	3.0875	6	2.9133	11	2.8953	2	3.1748	5	3.1572
43	3	3.0781	7	3.0609	1	3.5534	3	3.5344	2	3.1313	6	3.1120	7	2.9354	11	2.9189	2	3.1987	5	3.1820
44	3	3.1008	7	3.0847	1	3.5787	3	3.5614	2	3.1543	6	3.1361	7	2.9572	11	2.9420	2	3.2223	5	3.2064
45	3	3.1231	7	3.1081	1	3.6046	3	3.5868	3	3.1771	6	3.1598	7	2.9786	12	2.9573	2	3.2455	5	3.2304
46	3	3.1453	7	3.1311	1	3.6303	3	3.6129	3	3.1996	6	3.1831	7	2.9997	12	2.9801	2	3.2684	5	3.2540
47	4	3.1671	8	3.1457	1	3.6550	3	3.6388	3	3.2217	6	3.2060	7	3.0205	12	3.0024	2	3.2909	5	3.2773
48	4	3.1887	8	3.1684	1	3.6798	3	3.6637	3	3.2436	6	3.2286	7	3.0410	12	3.0243	2	3.3131	5	3.3002
49	4	3.2100	8	3.1906	1	3.7032	3	3.6887	3	3.2651	6	3.2509	7	3.0612	12	3.0458	2	3.3351	6	3.3137
50	4	3.2309	8	3.2126	1	3.7286	3	3.7123	3	3.2864	6	3.2728	7	3.0811	12	3.0670	2	3.3567	6	3.3363
51	4	3.2515	8	3.2342	1	3.7528	3	3.7383	3	3.3074	6	3.2944	8	3.1007	13	3.0814	2	3.3781	6	3.3585
52	4	3.2720	8	3.2555	1	3.7757	3	3.7619	3	3.3281	7	3.3078	8	3.1202	13	3.1022	2	3.3992	6	3.3805
53	4	3.2922	8	3.2764	1	3.7969	3	3.7851	3	3.3485	7	3.3291	8	3.1394	13	3.1227	2	3.4200	6	3.4021
54	4	3.3122	8	3.2972	1	3.8199	3	3.8067	3	3.3687	7	3.3501	8	3.1584	13	3.1429	2	3.4405	6	3.4234
55	4	3.3318	8	3.3177	1	3.8414	3	3.8296	3	3.3887	7	3.3709	8	3.1772	13	3.1627	2	3.4608	6	3.4445
56	4	3.3513	8	3.3378	1	3.8641	3	3.8513	3	3.4084	7	3.3913	8	3.1957	13	3.1823	2	3.4809	6	3.4652
57	4	3.3705	8	3.3577	1	3.8855	3	3.8741	3	3.4278	7	3.4115	8	3.2140	14	3.1959	2	3.5007	6	3.4857
58	4	3.3894	9	3.3709	1	3.9075	3	3.8957	3	3.4471	7	3.4314	8	3.2320	14	3.2151	2	3.5203	6	3.5060
59	4	3.4082	9	3.3904	1	3.9289	3	3.9177	3	3.4661	7	3.4511	8	3.2499	14	3.2341	3	3.5398	6	3.5260
60	4	3.4267	9	3.4098	1	3.9502	3	3.9402	3	3.4849	7	3.4705	9	3.2675	14	3.2529	3	3.5590	6	3.5458
61	4	3.4450	9	3.4289	1	3.9700	4	3.9502	3	3.5035	7	3.4897	9	3.2851	14	3.2714	3	3.5781	6	3.5653
62	4	3.4634	9	3.4480	1	3.9911	4	3.9704	3	3.5219	7	3.5087	9	3.3024	14	3.2896	3	3.5969	6	3.5846
63	4	3.4814	9	3.4667	1	4.0114	4	3.9918	3	3.5401	7	3.5275	9	3.3196	15	3.3025	3	3.6156	6	3.6037
64	4	3.4991	9	3.4852	1	4.0315	4	4.0123	3	3.5582	7	3.5460	9	3.3365	15	3.3205	3	3.6340	7	3.6156
65	5	3.5167	9	3.5034	1	4.0516	4	4.0327	3	3.5760	7	3.5644	9	3.3533	15	3.3382	3	3.6523	7	3.6345
66	5	3.5342	9	3.5214	1	4.0715	4	4.0530	3	3.5936	8	3.5763	9	3.3699	15	3.3558	3	3.6703	7	3.6531
67	5	3.5515	9	3.5393	1	4.0928	4	4.0731	3	3.6111	8	3.5945	9	3.3863	15	3.3731	3	3.6882	7	3.6716
68	5	3.5686	9	3.5571	1	4.1128	4	4.0947	3	3.6284	8	3.6124	9	3.4025	15	3.3902	3	3.7059	7	3.6898
69	5	3.5857	10	3.5690	1	4.1299	4	4.1137	3	3.6456	8	3.6301	10	3.4187	15	3.4071	3	3.7235	7	3.7079
70	5	3.6025	10	3.5866	1	4.1489	4	4.1321	3	3.6625	8	3.6477	10	3.4347	16	3.4193	3	3.7409	7	3.7258
71	5	3.6191	10	3.6039	1	4.1678	4	4.1513	4	3.6794	8	3.6651	10	3.4505	16	3.4361	3	3.7581	7	3.7434
72	5	3.6356	10	3.6210	1	4.1844	4	4.1703	4	3.6961	8	3.6823	10	3.4662	16	3.4526	3	3.7751	7	3.7610
73	5	3.6520	10	3.6379	1	4.2031	4	4.1869	4	3.7126	8	3.6993	10	3.4818	16	3.4689	3	3.7920	7	3.7783
74	5	3.6682	10	3.6547	1	4.2215	4	4.2058	4	3.7291	8	3.7161	10	3.4972	16	3.4850	3	3.8087	7	3.7955
75	5	3.6842	10	3.6713	1	4.2394	4	4.2243	4	3.7453	8	3.7328	10	3.5124	16	3.5010	3	3.8253	7	3.8125
76	5	3.7003	10	3.6877	1	4.2578	4	4.2429	4	3.7615	8	3.7494	10	3.5275	17	3.5127	3	3.8418	7	3.8294
77	5	3.7161	10	3.7042	1	4.2739	4	4.2609	4	3.7774	8	3.7658	10	3.5425	17	3.5285	3	3.8581	7	3.8460
78	5	3.7315	10	3.7203	1	4.2913	4	4.2768	4	3.7933	8	3.7820	11	3.5574	17	3.5441	3	3.8742	7	3.8626
79	5	3.7470	10	3.7360	1	4.3090	4	4.2948	4	3.8090	8	3.7981	11	3.5722	17	3.5596	3	3.8902	7	3.8790
80	5	3.7624	11	3.7472	1	4.3263	4	4.3122	4	3.8246	8	3.8140	11	3.5868	17	3.5749	3	3.9061	7	3.8952
81	5	3.7776	11	3.7631	1	4.3432	4	4.3295	4	3.8400	9	3.8249	11	3.6013	17	3.5901	3	3.9219	7	3.9114
82	5	3.7931	11	3.7788	1	4.3601	4	4.3466	4	3.8554	9	3.8406	11	3.6157	17	3.6051	3	3.9375	8	3.9220
83	5	3.8081	11	3.7947	1	4.3769	4	4.3635	4	3.8706	9	3.8563	11	3.6300	18	3.6163	3	3.9530	8	3.9379
84	6	3.8230	11	3.8101	1	4.3951	4	4.3802	4	3.8856	9	3.8718	11	3.6441	18	3.6311	3	3.9684	8	3.9537
85	6	3.8380	11	3.8254	1	4.4101	4	4.3986	4	3.9006	9	3.8872	11	3.6581	18	3.6458	3	3.9836	8	3.9694
86	6	3.8527	11	3.8407	1	4.4250	4	4.4136	4	3.9154	9	3.9024	11	3.6720	18	3.6604	3	3.9988	8	3.9850
87	6	3.8673	11	3.8557	1	4.4411	4	4.4286	4	3.9301	9	3.9176	12	3.6859	18	3.6748	3	4.0138	8	4.0004
88	6	3.8818	11	3.8706	1	4.4570	4	4.4446	4	3.9448	9	3.9326	12	3.6997	18	3.6891	3	4.0287	8	4.0157
89	6	3.8962	11	3.8855	1	4.4728	4	4.4605	4	3.9593	9	3.9474	12	3.7133	19	3.6999	3	4.0435	8	4.0309
90	6	3.9105	11	3.9002	1	4.4884	4	4.4763	4	3.9736	9	3.9622	12	3.7268	19	3.7140	3	4.0582	8	4.0459
91	6	3.9247	12	3.9107	1	4.5039	4	4.4919	4	3.9879	9	3.9768	12	3.7403	19	3.7281	3	4.0727	8	4.0609
92	6	3.9386	12	3.9253	1	4.5192	4	4.5074	4	4.0021	9	3.9914	12	3.7536	19	3.7420	4	4.0872	8	4.0757
93	6	3.9528	12	3.9396	1	4.5371	4	4.5231	4	4.0162	9	4.0058	12	3.7668	19	3.7558	4	4.1016	8	4.0904
94	6	3.9667	12	3.9541	1	4.5515	4	4.5406	4	4.0302	9	4.0201	12	3.7799	19	3.7695	4	4.1160	8	4.1050
95	6	3.9806	12	3.9683	1	4.5664	4	4.5549	4	4.0441	9	4.0343	12	3.7930	19	3.7831	4	4.1302	8	4.1195
96	6	3.9942	12	3.9826	1	4.5798	4	4.5698	4	4.0578	10	4.0442	13	3.8060	20	3.7934	4	4.1443	8	4.1339
97	6	4.0079	12	3.9966	1	4.5944	4	4.5831	4	4.0715	10	4.0583	13	3.8189	20	3.8068	4	4.1583	8	4.1482
98	6	4.0214	12	4.0105	1	4.6089	4	4.5976	5	4.0851	10	4.0723	13	3.8317	20	3.8202	4	4.1722	8	4.1623
99	6	4.0349	12	4.0244	1	4.6232	4	4.6121	5	4.0987	10	4.0861	13	3.8444	20	3.8334	4	4.1860	8	4.1764
100	6	4.0482	12	4.0381	1	4.6374	4	4.6263	5	4.1121	10	4.0999	13	3.8570	20	3.8465	4	4.1997	9	4.1860
\end{filecontents*}
\DTLloaddb[noheader,keys = {a,b,c,d,e,f,g,h,i,j,k,l,m,n,o,p,r,q,s,t,u},omitlines = 0]{data}{table_out.dat}
\setcounter{LTchunksize}{40}
\begin{longtable}[c]{|*{13}{c|}}
\caption{  Bondary harmonics and respective energy of effective spectral width for linear polarization components of SR and unpolarized radiation}\\ \hline
 $\Lambda$ & $\nu^{(1)}_{2}$ & $\gamma_{2}$ & $\nu^{(2)}_{2}$ & $\tilde\gamma_{2}$ & $\nu^{(1)}_{3}$ & $\gamma_{3}$ & $\nu^{(2)}_{3}$ & $\tilde\gamma_{3}$ & $\nu^{(1)}_{0}$ & $\gamma_{0}$ & $\nu^{(2)}_{0}$ & $\tilde\gamma_{0}$\\ \hline \endhead   \hline \endfoot
\DTLforeach[\value{DTLrowi}<101]{data}{ \a =a,\b=b,\c=c,\d=d,\e=e,\f=f,\g=g,\h=h,\i=i,\j=j,\k=k,\l=l,\m=m,\n=n,\o=o,\p=p,\r=r,\q=q,\s=s,\t=t,\u=u}{\DTLiffirstrow{}{\\ } \a & \b & \c & \d & \e &\f & \g & \h & \i & \j & \k & \l & \m  }
\end{longtable}

\begin{longtable}[c]{|*{9}{c|}}
\caption{  Bondary harmonics and respective energy of effective spectral width for circle polarization components  of SR}\\ \hline
  $\Lambda$& $\nu^{(1)}_{1}$&$\gamma_{1}$&$\nu^{(2)}_{1}$&$\tilde\gamma_{1}$  &$\nu^{(1)}_{-1}$&$\gamma_{-1}$&$\nu^{(2)}_{-1}$&$\tilde\gamma_{-1}$  \\ \hline \endhead   \hline \endfoot
\DTLforeach[\value{DTLrowi}<101]{data}{ \a =a,\b=b,\c=c,\d=d,\e=e,\f=f,\g=g,\h=h,\i=i,\j=j,\k=k,\l=l,\m=m,\n=n,\o=o,\p=p,\r=r,\q=q,\s=s,\t=t,\u=u}{\DTLiffirstrow{}{\\ } \a &  \n & \o & \p & \r &\q &\s &\t &\u  }
\end{longtable}

 \DTLdeletedb{data}
 
 \section*{Acknowledgements}
 The work of Bagrov and Gitman is partially supported by RFBR research project No. 15- 02-00293a and by Tomsk State University Competitiveness Improvement Program. Gitman thanks CNPq and FAPESP for their permanent support. Levin thanks CNPq for permanent support.  

\end{document}